\documentstyle[pre,aps,epsfig,multicol]{revtex}
\begin{document}
\draft
\title {Path-integral formulation of stochastic processes 
for exclusive particle systems}

\author{Su-Chan Park and Doochul Kim}

\address{School of Physics, Seoul National University, Seoul, Korea}

\author{Jeong-Man Park}

\address{Department of Physics, The Catholic University of Korea, Puchon, Korea}

\date{\today}
\maketitle
\begin{abstract}
We present a systematic formalism to derive a path-integral
formulation for hard-core particle systems far from
equilibrium. 
Writing the master equation for a stochastic
process of the system in terms of the annihilation and creation 
operators with mixed commutation relations,
we find the Kramers-Moyal coefficients for the corresponding
Fokker-Planck equation (FPE), and the stochastic differential 
equation (SDE) is
derived by connecting these coefficients in the FPE to
those in the SDE. Finally, the SDE is mapped onto a field theory
using the path integral, giving the field-theoretic action, which
may be analyzed by the renormalization group method. We apply this
formalism to a two-species reaction-diffusion system with 
drift, finding a universal decay 
expoent for the long-time behavior of the average
concentration of particles in arbitrary dimension.
\end{abstract}

\pacs{PACS number(s): 82.20.Db, 05.40.--a, 05.70.Ln, 82.20.Mj}

\begin{multicols}{2}
\narrowtext 
In recent years, nonequilibrium phenomena such as
nonequilibrium phase transitions, bifurcations, and synergetics
have attracted much attention\cite{P97}, not only because of their
connections to a variety of important physical problems (pattern
formation, morphogenesis, self-organization, etc.), but also
because of the analytic challenge due to lack of a general
formalism for nonequilibrium systems, in contrast to equilibrium
statistical mechanics, which has well-established concepts and tools. In
pursuit of a general formalism, statistical physicists have 
investigated nonequilibrium phase transitions in
lattice models over the last decade\cite{MD99}. As lattice models have played
a central role in equilibrium statistical mechanics, they will also be
important in nonequilibrium statistical mechanics.
In particular, theoretical analysis of reaction-diffusion systems
where both diffusion and reaction take place on the lattice is
relevant to the understanding of a wide class of nonequilibrium
phenomena in nature\cite{MG98}. It has long been recognized that the
mean-field rate equations are not applicable to reaction-diffusion
systems in low dimensions. 
After Doi and others introduced the field-theoretic method
using the bosonic coherent state path integral\cite{Doi}, Lee
and Cardy using the renormalization group (RG) approach
have improved on this method\cite{L94,LC95} in the description 
of the anomalous kinetics in these systems.
%it is only recently that Lee and Cardy,
%using the renormalization group (RG) 
%approach\cite{L94,LC95}
%have improved on the field theoretic method which was
%formulated by Doi and others using the bosonic coherent state
%path integral\cite{Doi} in the description of the anomalous kinetics
%in these systems.
Assuming the systems are in the
low density regime, Lee and Cardy rewrite the master equation for
the Markov process as the bosonic {\it Hamiltonian
}. The Hamiltonian in turn can be
mapped onto field theory and analyzed by the renormalization group
method in arbitrary dimensions. For simple models such as
$A+A\rightarrow\emptyset$ and $A+B\rightarrow\emptyset$, this bosonic field-thoeretic method
provided the correct time dependence for the density decay in low
dimensions\cite{L94,LC95,DP98,PD98}.

Despite the successes achieved by the bosonic field theory for
reaction-diffusion systems, there are still many open problems.
Driven reaction-diffusion systems\cite{J95,IKR95}, multi-species
adsorption models\cite{BB97}, and epidemic models are some examples to
which the bosonic field theory cannot be applied since the
steady states of these systems cannot be assumed to be in  a low
density regime. In these systems, the hard-core property of
the particles is important and the bosonic approach fails. In response
to these challenges, there have been many attempts to take the
hard-core property into account. Brunel {\it et al}.\cite{BOvW99} and Bares and
Mobilia\cite{BM99} formulated fermionic field theories for a single-species
reaction-diffusion process confined to one space dimension.
However, these fermionic field theories are very hard to extend in practice to 
higher spatial dimensions or to multispecies processes.

We have focused on the extensibility of the field
theory to multispecies processes and to higher spatial
dimensions including the hard-core exclusion property of particles. In
this paper, we present a systematic formalism to derive the field
theory for hard-core particles and apply this method to a
two-species driven reaction-diffusion (DRD) system in arbitrary
spatial dimension. In the two-species DRD system, each particle
attempts moves to the right and to the left with different
hopping rates, and the attempt is successful only if the particle
lands on an unoccupied site. If the particle lands on a site
occupied by a same-species particle, the hopping attempt is
rejected, but if it lands on a site occupied by an
opposite-species particle, the $A+B\rightarrow \emptyset$ reaction occurs and
both particles disappear. For this system, one might expect the
long-time kinetics to be the same as that of $A+B\rightarrow \emptyset$ with
isotropic diffusion, by a Galilean transformation, and the density
should decay in time as $t^{-d/4}$ for $d \le 4$ and as $t^{-1}$
for $d \ge 4$. However, some extensive numerical
simulations by Janowsky\cite{J95}
and Ispolatov {\it et al.}\cite{IKR95} indicate that the density 
decays as $t^{-1/3}$ asymptotically in one dimension, and others by
ben-Avraham {\it et al.}\cite{bA95} are inconclusive 
concerning the exponent of the density decay.
Consequently, to study this system analytically, the hard-core
property of the particles should be incorporated properly 
into the field theory.  Our general formalism provides a systematic method to
derive the field theory for this system and with the application
of the renormalization group derives the long-time behavior
as predicted by Janowsky and Ispolatov {\it et al.} for
density decay as $t^{-1/3}$ in one dimension.

In general, the dynamics of a stochastic particle system is described 
by a master equation governing the time evolution of the probability
$P({\cal C};t)$ for the system to be in a given microscopic 
configuration ${\cal C}$ at time $t$. 
For a multispecies reaction-diffusion system with hard-core particles,
the microscopic configuration ${\cal C}$ is represented by the set of the
particle numbers of each species at each lattice site; ${\cal C} = \{ n_i^\alpha
\}$ where the greek index $\alpha$ stands for the particle species,
the latin index $i$ runs over all lattice sites 
in arbitrary spatial dimension,
and $n_i^\alpha$ is restricted to $0$ or $1$.
Introducing the annihilation and creation
operators satisfying the mixed commutation relations
\begin{equation}
\{a_i^\alpha, a_i^{\alpha\dag}\} = 1 -
 \sum_{\gamma \neq \alpha} a_i^{\gamma\dag} a_i^\gamma,
\;\;\; {a_i^\alpha}{a_i^\beta} = {a_i^{\alpha\dag}}{a_i^{\beta\dag}}  = 0,
\end{equation}
\begin{equation}
[{a_i^\alpha},{a_j^{\beta\dag}}]=[{a_i^\alpha},{a_j^\beta}]=
[{a_i^{\alpha\dag}},{a_j^{\beta\dag}}]=0 \; {\rm for} \; i \neq j,
\end{equation}
and defining the state vector 
$|\Psi ;t\rangle \equiv \sum_{\cal C} P({\cal C};t) |{\cal C} \rangle$, 
the master equation can be written as a Schr\"odinger-like equation\cite{K72},
\begin{equation}
\frac{\partial}{\partial t} |\Psi;t\rangle = - {\cal H}|\Psi;t\rangle,
\end{equation}
where $\cal H$ is an evolution operator, often called a 
Hamiltonian, expressed in terms of
$a$'s and $a^{\dagger}$'s.
The formal solution for the initial condition $|\Psi;0\rangle$ is, 
straightforwardly, $|\Psi ;t\rangle = e^{- {\cal H} t} |\Psi ;0\rangle$, and
the average of any quantity $f$ may be expressed as
\begin{equation}
\langle f(t) \rangle \equiv \sum_{\{n_i^\alpha\}}
 f(\{n_i^\alpha\}) P(\{n_i^\alpha\};t) =
\langle \cdot | \hat f e^{-{\cal H} t} |\Psi ;0\rangle,
\label{average}
\end{equation}
where $\langle \cdot | $ is the projection state defined 
as the sum of all possible microscopic states,
{\it i.e.}, $ \langle \cdot |
\equiv \sum_{\{n_i^\alpha\}} \langle \{n_i^\alpha\}|$.
For a given observable $f(\{n_i^\alpha\})$, we find the corresponding operator $\hat{f}$
by replacing the variables $n_i^\alpha$ by the operator $a_i^{\alpha \dag}
a_i^\alpha$.
In what follows, we shall be mainly interested in averages of particle
numbers($\hat f = a_i^{\alpha\dag} a_i^\alpha$) at site $i$ and
their two-point correlation functions ($\hat f = a_i^{\alpha\dag} a_i^\alpha
a_j^{\beta\dag} a_j^\beta$). The time derivative of Eq.(\ref{average})
is formally found to be
\begin{eqnarray}
{d \over dt} \langle f(t) \rangle =-\langle \cdot|\hat f{\cal H}|\Psi \rangle_t
= \langle [{\cal H},\hat f] \rangle,
\label{time_deriv_of_quantity}
\end{eqnarray}
where we used the probability conservation condition $\langle \cdot | 
{\cal H} = 0$.
Since the Hamiltonian describes a stochastic process, in general $\cal H$ is not
Hermitian. Thus, $[{\cal H}, \hat f]$ will have creation and annihilation
operators that do not form number operators. However, the projection state
$\langle \cdot |$ acting on $[{\cal H}, \hat f]$ makes it possible to express
the right-hand side of Eq. (\ref{time_deriv_of_quantity}) only
with number operators.
Using the identity from 
the property of the projection state\cite{PPK} 
\begin{equation} \label{id}
\langle \cdot | \left (a_i^{\beta\dag}+ \sum_{\alpha } a_i^{\alpha} \right )
=\langle \cdot | 
\end{equation}
for any $\beta$, we 
eliminate all the creation operators in Eq. (\ref{time_deriv_of_quantity}),
and any annihilation operator can be interpreted as a number operator
because
$
\langle \cdot | a_i^{\beta\dag} a_i^\beta =
 \langle \cdot | (1 - \sum_\alpha a_i^\alpha) a_i^\beta
= \langle \cdot | a_i^\beta.
$

Since the Kramers-Moyal coefficients 
$C_i^\alpha$, $C_{ij}^{\alpha \beta}$ in the Fokker-Planck equation
\begin{eqnarray}
\label{fokplan}
{\partial P \over \partial t}
= - {\partial \over \partial \rho_i^\alpha} [C^\alpha_i(\{\rho\}) P ]
+ {1\over 2}
{\partial^2 \over \partial \rho_i^\alpha \partial \rho_j^\beta}
 [ C_{ij}^{\alpha\beta}(\{\rho\}) P ]
\end{eqnarray}
are related to the time evolution of the one-point and two-point 
correlation functions of the number operator
\begin{equation}
{d \over dt} \langle \rho_i^\alpha \rangle = \langle  C^\alpha_i
\rangle,~~~
{d \over dt}\langle \rho_i^\alpha \rho_j^\beta\rangle = \langle
\rho_i^\alpha C_j^\beta  + \rho_j^\beta C_i^\alpha +   C_{ij}^{\alpha\beta}
\rangle ,
\end{equation}
we find the Kramers-Moyal coefficients in terms of
the annihilation and creation operators\cite{PPK}
\begin{equation}
\langle C_i^\alpha \rangle = 
\langle [{\cal H}, a_i^{\alpha\dag} a_i^\alpha ] \rangle, \;\;\;
\langle C_{ij}^{\alpha\beta} \rangle = \langle [
 a_i^{\alpha \dag} a_i^\alpha,
[{\cal H}, a_j^{\dag\beta} a_j^\beta] ] \rangle,
\end{equation}
by interpreting the number operator as a density of particles.

Next we consider how we write down the stochastic differential equation
when the Fokker-Planck equation is known. Recalling the reverse problem,
a stochastic differential equation
\begin{equation}
\label{gen_lan}
\dot {\rho_i^\alpha} = h_i^\alpha(\{\rho\}) + g_{ij}^{\alpha\beta}
(\{\rho\}) \xi_j^\beta(t)
\end{equation}
with $
\langle \xi^\alpha_i(t) \xi^\beta_j(t^\prime) \rangle =  \delta^{\alpha
\beta}\delta_{ij}
\delta(t - t ^\prime)$ can be connected to the Fokker-Planck equation
with the coefficient functions
$
C_i^\alpha(\{\rho\}) = h_i^\alpha$ and $C_{ij}^{\alpha\beta}
(\{\rho\}) = g^{\alpha\gamma}_{ik}g^{\beta\gamma}_{jk}
$
in the It\^o interpretation\cite{vK97}.

Representing the stochastic differential equation in the path-integral
formulation, the generating
functional $Z$ of correlation functions can be written as 
$Z = \int {\cal D} \rho {\cal D} \tilde\rho e^{-S}$ with the action\cite{MSR73}
\begin{eqnarray}
S = \int dt \left (
 \tilde \rho_{i}(t)
 ( \partial_t \rho_{i}(t)
- C_{i}) - {1 \over 2}
\tilde \rho_{i}(t) \tilde \rho_{j}(t) C_{ij} \right ).
\label{action}
\end{eqnarray}
The response field $\tilde \rho$ has been introduced as the conjugate field to
the Langevin force.
After performing a suitable continuum limit for
the action, we obtain the continuum field description for
microscopic discrete models.
Thus, by mapping the stochastic differential equation derived
from the Fokker-Planck equation into the path-integral formalism,
 we obtain a field-theoretic action
describing the stochastic process, which in turn may be examined by RG analysis.

Now we apply our formalism to reaction-diffusion systems.
As the paradigmatic example, we consider the asymmetric diffusion process
with
$A + B \rightarrow \emptyset$ reaction in $d$-dimensional space.
The diffusion constant for an $A~(B)$ particle is $D^A~(D^B)$  and
along the direction of the driving force (say, the ``parallel'' direction) the
diffusion is asymmetric with the drift rate
$v_A/2~(v_B/2)$ for an $A~(B)$ particle.
The reaction occurs with rate $\lambda/2d$
when
two different species occupy the nearest neighbor sites in a $d$ dimensional
hypercubic lattice.
The Hamiltonian generating the time evolution of the system is found to be
${\cal H} = \sum_{\vec n} \left [ {\cal H}^{\rm dif}_{\vec n} +
{\cal H}^{\rm dr}_{\vec n} + {\cal H}^{\rm re}_{\vec n}
\right ] $ with ($\vec e_i$ is the unit vector along the direction $i$)
\begin{eqnarray}
{\cal H}^{\rm dif}_{ \vec n } &=& - \sum_{i=1}^d
\left[ D^A \left(a_{ \vec n } a_{ \vec n +\vec e_i }^\dag
+ a_{ \vec n }^\dag a_{ \vec n + \vec e_i } \right) 
\right. \nonumber \\ 
&& \left. 
+ D^B \left( b_{ \vec n } b_{ \vec n +\vec e_{i} }^\dag 
+ b_{ \vec n }^\dag b_{ \vec n +\vec e_{i} } \right)
\right],\nonumber \\
{\cal H}^{\rm dr}_{ \vec n } &=& 
- { v_A \over 2 } \left( a_{ \vec n } a_{ \vec n +\vec e_{\|} }^\dag
- a_{ \vec n }^\dag a_{ \vec n + \vec e_{\|} } \right)
\nonumber \\ 
& & \hspace{1.5cm}
 - { v_B \over 2 } \left( b_{ \vec n } b_{ \vec n + \vec e_{\|} }^\dag
- b_{ \vec n }^\dag b_{ \vec n + \vec e_{\|} } \right),\\
{\cal H}^{\rm re}_{ \vec n } &=& - {\lambda \over 2d } \sum_{i=1}^d 
\left( a_{ \vec n } b_{ \vec n + \vec e_{i} }
+ b_{ \vec n } a_{ \vec n + \vec e_{i} } \right),\nonumber
\end{eqnarray}
where we left out the diagonal terms because they give no
contribution to the commutation relations.
Following the steps given above, we find the field-theoretic action
for the system after taking the continuum limit.
\begin{eqnarray}
S &=& \int dt d{\vec x} \Biggl ( \tilde{a} (\partial_t - D^A
\nabla^2 ) a
+ \tilde{b} (\partial_t - D^B\nabla^2  ) b  \nonumber \\
&-& 2v_A a(\rho_{m}-a-b) \partial_\| \tilde{a}
- 2v_B b(\rho_{m}-a-b) \partial_\| \tilde{b} \nonumber \\
&+& M^A ( \nabla\tilde{a} )^{2}  a(\rho_{m}-a-b)+
M^B ( \nabla\tilde{b} )^{2}  b(\rho_{m}-a-b)
\nonumber \\
&+& {\lambda \over 2} \left[ 2(\tilde{a} + \tilde{b}) 
- (\tilde{a} + \tilde{b})^2 \right] ab \Biggr ),
\end{eqnarray}
in terms of the density fields $(a,b)$ of each species and the conjugate
response fields $(\tilde{a},\tilde{b})$. 
The hard-core property of particles is manifest in the action and $\rho_{m}$ 
is the density cutoff due to the hard-core property.
Since the densities are restricted to $a,b \ge 0$, we shift the fields by
$\alpha = 2 a - \rho_m$, $\beta  = 2 b - \rho_m $, $\tilde{\alpha}=\tilde{a}-1$, and
$\tilde{\beta}=\tilde{b}-1$, in order to apply a perturbative RG analysis.
Skipping all the irrelevant terms, we get the reduced action
\begin{eqnarray}
S &=& \int dt d{\vec x} \Biggl( \tilde\alpha (\partial_t - D
\nabla^2 ) \alpha
+ \tilde\beta (\partial_t - D\nabla^2  )\beta \nonumber
\\
&-& {v \over2}\alpha^2\partial_\| \tilde\alpha
- {v \over2}\beta^2\partial_\| \tilde\beta
+ M \tilde\alpha \nabla^2 \tilde\alpha + M \tilde\beta \nabla^2 \tilde\beta
\nonumber \\
&-& {\lambda \over 2} \left [ (\tilde\alpha + \tilde\beta)^{2} + 2 (\tilde\alpha + \tilde\beta) \right ]
(\rho_m + \alpha ) (\rho_m + \beta) \Biggr )
\end{eqnarray}
in the case of $D^A = D^B = D$, $v_A=v_B=v$, and $M^A=M^B=M$ with 
$D(M)\nabla^2 = D_{\|}(M_{\|})\nabla_{\|}^2 + 
D_{\perp}(M_{\perp}) \nabla_{\perp}^2$.
From power counting with shifted fields, we find the upper critical
dimension $d_c = 2$. The scaling dimension of the coupling constant $v$
indicates that the drift term is effective only for fewer than two dimensions, 
and for
$d \ge 2$ the action becomes equivalent to the action derived by
Lee and Cardy using the bosonic approach for the symmetric
reaction-diffusion system $A+B \rightarrow\emptyset$ without drift.

\begin{figure}
\epsfig{figure=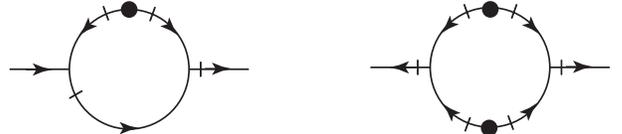,width=8cm}
\caption{The one-loop diagrams contributing to the renormalization of (left)
$D_{\|}$ and (right) $M_{\|}$. The legs with the outgoing arrow are for the 
response fields $(\tilde{\alpha},\tilde{\beta})$ and the legs with the 
incoming arrow for the density fields $(\alpha,\beta)$.
The bar denotes spatical differentiation and the dot
an $M$ vertex.
}
\label{DM}
\end{figure}

\begin{figure}
\epsfig{figure=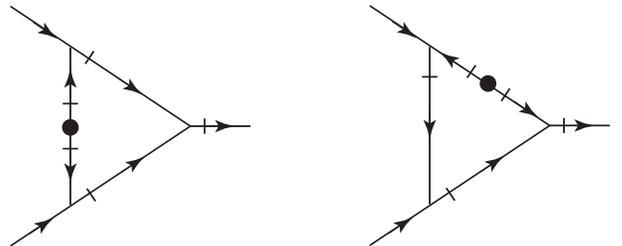,width=8cm}
\caption{The one-loop diagrams contributing to the renormalization of
$v$. These two diagrams cancel each other.
}
\label{vv}
\end{figure}

We use the Wilson RG method to analyze the long-time kinetics of
the action. The flow equations in $d = 2 - \varepsilon$ dimensions,
to one loop-order, are 
\begin{eqnarray} 
{dD_{\perp} \over dl}&=& (z-2)D_{\perp} ,~~~
{dM_{\perp} \over dl} = (z-2)M_{\perp}, \nonumber\\
{dD_{\|} \over dl} &=& (z-2)D_{\|} + \frac{D_{\|}}{8}
\left( 3+\frac{M_{\|}}{D_{\|}}\right) g, \nonumber \\ 
{dM_{\|} \over dl} &=& (z-2)M_{\|} + \frac{M_{\|}}{16}
\left( 3\frac{M_{\|}}{D_{\|}} + 2 + 3\frac{D_{\|}}{M_{\|}} \right), \\ 
{d\lambda \over dl} &=& (z-d)\lambda + \frac{\lambda^2}{4\pi \sqrt{
D_{\|}D_{\perp}}} ,~~~
{dv \over dl } = \left( z-1-\frac{d}{2} \right)v, \nonumber 
\end{eqnarray} 
where $ g = v^{2}/4\pi D_{\|}^{3/2}D_{\perp}^{1/2}$. 
The Feynman
diagrams that contribute to these equations are shown in Figs.~\ref{DM} 
and \ref{vv}. 

The dynamical exponent is given by $z = 2$, leaving $D_{\perp}$ and
$M_{\perp}$ unchanged under the RG flow. 
The flow equations for $D_{\|}$ and $M_{\|}$ have the same contribution  
and the ratio $D_{\|}/M_{\|}$ remains constant $(=1)$.
The reaction rate $\lambda$ is renormalized  only by the $\lambda$
terms, not the drift term. 
Combining the flow equations for $D_{\perp}$, $D_{\|}$, and $v$,
we find the flow equation for the expansion parameter $g$:
\begin{equation}
\frac{d\ln g}{dl} = (2-d) - \frac{3}{4}g.
\end{equation}
\noindent
For $d>d_c = 2$ we find
an infrared stable fixed point $g^* = 0$, and in a region of
attraction, $D_{\|}$ and $M_{\|}$ remain constant. The scaling
form of the average concentration of $A$ and $B$ particles [$c(t)
\equiv c_A(t) = c_B(t)$]\cite{DP98}
\begin{equation}
c(t) = e^{- d \ell} \left (
{n_0(\ell) \over 8 \pi^2 D_{\|}(\ell) ^{1/2} D_{\perp}(\ell) ^{(d-1)/2}
 t(\ell)^{d/2} } \right )^{1/2} ,
\end{equation}
 gives $c(t) \sim t^{-d /4}$ using the time flow equation
$t(\ell) = t e^{-\int_0^\ell z(\ell) d \ell}$. Below the critical
dimension $d_c$, there exists a nontrivial infrared stable fixed
point at $g^* = { 4 \over 3} \varepsilon$, and near this point
$D_{\|}$ and $M_{\|}$ flow as $e^{2\epsilon \ell/3}$. Thus, the average
concentration behaves as
\begin{equation}
c(t) \sim t^{- (d + 1)/6} .
\end{equation}

In summary, we have presented a systematic formalism to derive
the field-theoretic action for systems of hard-core particles.
Starting from the master equation for a stochastic process of the
system, we have constructed the Fokker-Planck equation by
introducing annihilation and creation operators with mixed
commutation relations. This Fokker-Planck equation is connected to
the stochastic differential equation by identifying the
coefficient functions in the It\^o interpretation. Finally, the
stochastic differential equation is mapped onto field theory
using the path integral, giving the field-theoretic action to be
analyzed by the RG method.

Although there have been many attempts to incorporate the
hard-core property of particles into field theory, our
formalism has a very important advantage over other attempts. Our
formalism can be applied to  multispecies systems in
arbitrary spatial dimension. As a paradigmatic example, we have
applied our formalism to the $A+B \rightarrow\emptyset$ reaction-diffusion
system with drift. Following straightforward steps to
obtain the action and applying the momentum-shell RG method, we
have calculated the long-time behavior for the average
concentration of particles. Power counting shows that the upper
critical dimension is $d_c = 2$, and the drift term affects the RG
flow only for fewer than two dimensions. Thus, for $d \ge 2$, the
hard-core action behaves the same as the bosonic action derived by
Lee and Cardy. The average concentration behaves as $t^{-1}$ for
$d \ge 4$ and  $t^{-d/4}$ for $ 4 \ge d \ge 2$ in the long-time
limits. Below the critical dimension, the drift term moves the
stable fixed point to the non-trivial one and the average
concentration behaves as $t^{-(d+1)/6}$ for $d \le 2$. These
results agree with the simulation results by Janowsky\cite{J95} and the
scaling arguments by Ispolatov {\it et al}\cite{IKR95}.

As mentioned before, our formalism has merit in 
extension to  multispecies and to higher spatial
dimensions. Also, it is necessary to use this formalism, not the
bosonic formalism, when the system has nonvanishing
concentrations in the steady states. The three-species
reaction-diffusion system\cite{BB97} and some other systems having
nonvanishing steady states are under investigation using this
formalism.

\vspace{0.3cm}
This research was supported by the KOSEF through Grand No. 
981-0202-008-2.

\end{multicols}
\end{document}